\documentclass[letter]{aa}

\usepackage{graphicx}
\usepackage{txfonts}
\usepackage{color}

\begin{document}

   \title{Color profiles of disk galaxies at $z=1$--$3$ observed with JWST: Implications for outer-disk formation histories}

   \author{Si-Yue Yu\inst{\ref{ipmu}}, 
          Dewang Xu\inst{\ref{PKU}},  
          Boris S. Kalita\inst{\ref{ipmu},\ref{kiaa},\ref{CD3}},
          Sijia Li\inst{\ref{xmu},\ref{ucas}},  
          John D. Silverman\inst{\ref{ipmu},\ref{CD3},\ref{UTokyo}}
          Xinyue Liang\inst{\ref{xmu}},  
          \and 
          Taotao Fang\inst{\ref{xmu}}, 
          }

   \institute{ 
       Kavli Institute for the Physics and Mathematics of the Universe (Kavli IPMU, WPI), UTIAS, Tokyo Institutes for Advanced Study, University of Tokyo, Chiba, 277-8583, Japan \\ \email{si-yue.yu@ipmu.jp, phyyueyu@gmail.com} \label{ipmu}
        \and 
   Department of Astronomy, Peking University, 5 Yiheyuan Road, Haidian District, Beijing, 100871, China \\ \email{dwxu.astro@gmail.com} \label{PKU}
   \and
   Kavli Institute for Astronomy and Astrophysics, Peking University, Beijing 100871, People's Republic of China \label{kiaa}
   \and 
   Center for Data-Driven Discovery, Kavli IPMU (WPI), UTIAS, The University of Tokyo, Kashiwa, Chiba 277-8583, Japan \label{CD3}
   \and 
   Department of Astronomy, Xiamen University, Xiamen, Fujian 361005, People's Republic of China \label{xmu}
   \and 
   School of Astronomy and Space Science, University of Chinese Academy of Sciences (UCAS), Beijing 100049, China \label{ucas}
   \and 
   Department of Astronomy, School of Science, The University of Tokyo, 7-3-1 Hongo, Bunkyo, Tokyo 113-0033, Japan \label{UTokyo}
            }


\abstract{We investigate the deconvolved color profiles of 223 disk galaxies at redshifts of  $z=1$--3 observed by the James Webb Space Telescope (JWST) as part of the Cosmic Evolution Early Release Science survey (CEERS). The filters were selected to approximate the rest-frame $B-Y$ color, which is used to identify U-shaped color profiles ---those becoming progressively bluer with increasing radius, then turning redder beyond a specific point. We find that 36\% of Type~II (down-bending) disks exhibit U-shaped color profiles with a minimum at or near the disk break. In contrast, no Type~I (single-exponential) disks and only 9\% of Type~III (up-bending) disks show such a profile. The presence of U-shaped color profiles in Type~II disks likely arises from the interplay between a star-formation threshold and spiral- or bar-driven secular radial migration of older stars outward. The fraction of Type~II disks exhibiting a U-shaped color profile remains almost consistent across two redshift bins, $z=1$--2 and $z=2$--3, but is significantly lower than that observed in the local Universe, likely because the secular process of radial migration at high redshift may not have had sufficient time to significantly influence the disk structure. The absence of U-shaped color profiles in Type~II disks could point to rapid rather than secular radial star migration potentially caused by violent clump instabilities, transporting both younger and older stars to the outer disk. Our results provide useful constraints on the formation and evolution models of disk galaxies in the early Universe. 
    }


   \authorrunning{Yu et al.}
   \maketitle
   

\section{Introduction}

Since the seminal work by \cite{vanderKruit1979}, it has been established that the radial-surface-brightness profile of disk galaxies does not always decline exponentially but instead exhibits a sharp change after several radial scale lengths. \cite{pohlen2006} and \cite{Erwin2008} identified three main categories of surface brightness profiles: (1) Type~I, classical single exponential; (2) Type~II, down-bending double exponential; and (3) Type~III, up-bending double exponential. The transition between the two exponential segments is referred to as the disk break. In the local Universe, approximately 20\%, 50\%, and 30\% of disk galaxies observed in the optical display Type~I, Type~II, and Type~III surface-brightness profiles, respectively \citep{pohlen2006, Erwin2008, Gutirrez2011}. Consistent fractions have been obtained in the near-infrared  \citep{Laine_2014}. These fractions remain largely unchanged at intermediate redshifts ($z=0.1$--1.1) \citep{Azzollini2008}.  Leveraging the superior near-infrared imaging capabilities of JWST NIRCam, \cite{XuYu2024} study high-redshift disk galaxies observed in the Cosmic Evolution Early Release Science survey (CEERS) (PI: Finkelstein, ID=1345, \citealt{Finkelstein:2022}) and report fractions of 12.6\%, 56.7\%, and 34.8\% for the three types at $z=1$--3, respectively.

Simulations by \cite{Roskar2008} suggested that the Type~II disks result from a combination of star formation cut-off in the outer disk and secular radial star migration, and predicted a U-shape in their color profiles. Consistently, observations at redshift $z<1$ reveal that most Type~II disks exhibit a U-shaped color profile, with the minimum typically located at or near the disk break \citep{Azzollini2008_color, Bakos2008, Zheng2015, Marino2016}. In contrast, Type~I disks, possibly formed via protogalactic cloud collapse \citep[e.g.,][]{Freeman1970} or viscous redistribution \citep[e.g.,][]{Ferguson2001}, and Type~III disks, possibly formed via interaction with intergalactic environment \citep[e.g.,][]{Younger2007}, rarely show U-shaped color profiles.

At high redshifts $z\gtrsim1$, as the timescale for secular processes to exert significant influence is several gigayears \citep{Roskar2008}, spiral- or bar-driven radial migration may not have had sufficient time to make Type~II disks. Additionally, the evolution of bar fractions over cosmic time \citep{Kraljic2012, Rosas-Guevara2022} and the potential underdevelopment of spiral structures due to turbulence \citep{ElmegreenElmegreen2014} may further limit the impact of secular radial migration during this epoch.  These considerations suggest that the fraction of Type~II disks at high redshift might decrease; however, this fraction remains comparable to that in the local Universe \citep{XuYu2024}. Therefore, it is suggested that, alongside secular radial migration, a much more rapid mechanism ---likely driven by clump instabilities \citep{Bournaud2007, Bournaud2016}--- could be responsible for the formation of Type~II disks at high redshift. Such rapid processes do not necessarily lead to U-shaped color profiles.  This dichotomy between secular and rapid mechanisms raises questions about the formation mechanism of the high-redshift Type~II disks.

Studying the color profiles of high-redshift disks could significantly enhance our understanding of the physical processes that dominate disk evolution in the early Universe. However, the color profiles of these high-redshift disks remain unexplored. In this letter, we investigate the color profiles of disk galaxies at redshifts of $z=1$–3 using the JWST sample from \cite{XuYu2024} and discuss their implications for the formation and evolution of disk galaxies.

\section{Observation material} \label{OM}

\cite{XuYu2024} define a sample of 247 face-on disk galaxies at redshifts of $z=1$--3 from CEERS, excluding those with tidal features and selecting galaxies with an inclination of $i\leq60^\circ$, stellar mass of $M_* > 10^{10} \, \text{M}_\odot$, and a half-light radius of $R_e > 2\times \text{F356W full width at half maximum (FWHM)}$.  Although the sample may still include galaxies that have undergone minor mergers in the past, these events typically produce Type~III profiles rather than Type~II. Our analysis focuses on the color profiles of Type~II disks, reducing concerns over the potential effects of past minor mergers. The size criterion ensures robust extraction of deconvolved surface-brightness profiles, with the point spread function (PSF) effects removed.  Eliminating the PSF smoothing effect is crucial to study galaxy morphology \citep{Yu2023, Liang2024}.  The classification into Type~I, II, and III disks was based on the exponential-function fitting to the deconvolved F356W-band profiles \citep{XuYu2024}. The deconvolution process involves subtracting a PSF-convolved multi-Gaussian model from the original image and subsequently adding the model without PSF convolution back to the residual; for details, see \cite{XuYu2024}. Adopting this method, we extracted the F115W- and F150W-band surface-brightness profiles that are not available in \cite{XuYu2024}.  The extraction robustness is automatically ensured due to the narrower PSFs of the F115W- and F150W-band images. Galaxies lacking these two-band images were excluded, leaving 223 galaxies. In these galaxies, there are 41 Type~I, 135 Type~II, and 83 Type~III profiles. Some galaxies exhibit both Type II and Type III breaks. Scale lengths, $h_s$, and break radii, $R_{\rm break}$, were adopted from \cite{XuYu2024}.

Our construction of the color profiles to study these disks is based on the principle that one band traces young stars and another traces older stars. Consequently, we selected F115W or F150W as the blue band, and F356W as the red band. We avoid using F444W, as this would significantly reduce the number of galaxies due to the wider F444W PSF and the requirement of $R_e > 2 \times \text{FWHM}$ for robust deconvolution.  To facilitate optimal comparisons of color profiles across different redshift ranges, we selected filters that closely approximate rest-frame $B-Y$: for $z=1$--2, we used ${\rm F115W}-{\rm F356W}$, and for $z=2$--3 we used ${\rm F150W}-{\rm F356W}$. To analyze the radial color gradient, we first normalized the profile radius by the characteristic radius, which is set as $2\times h_s$ for Type~I disks and $R_{\rm break}$ for Type~II and III disks.  We then fitted linear functions to the color profile on both sides of the characteristic radius. The fitting regions are selected as the nearly monotonically increasing or decreasing intervals of data. The U-shape in the color profile is identified if the slope of the best-fit linear function at the low-radius side is less than zero, while that at the high radius side is greater than zero. The U-shape position is estimated as the radius where the two best-fit functions intersect.

\begin{figure}
        \centering
        \includegraphics[width=0.45\textwidth]{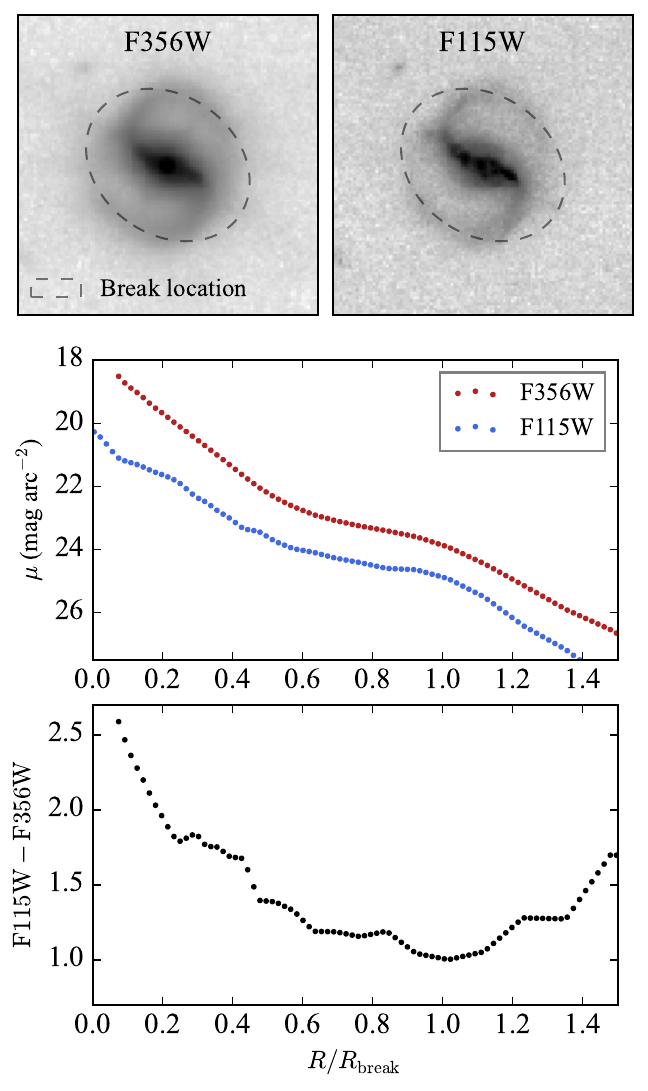}
        \caption{
        Illustration of the calculation of color profiles. The upper panels display the F356W- and F115W-band images of the CEERS galaxy at ${\rm RA}=215\fdg0922699$ and ${\rm Dec.}=52\fdg9221344$. The dashed ellipse marks the break location. The middle panel shows the extracted deconvolved surface-brightness profiles for F115W band, marked in blue, and for F356W band, marked in red. The calculated color profile, exhibiting a clear U-shape, is shown in the bottom panel.
        }
        \label{example}
\end{figure}

Figure \ref{example} presents an example of color-profile extraction for a CEERS galaxy at ${\rm RA}=215\fdg0922699$ and ${\rm Dec.}=52\fdg9221344$. The upper panels display the F356W- and F115W-band images with the dashed ellipse marking the break location, while the middle panel shows the extracted deconvolved surface brightness profiles in F356W (red) and F115W (blue). The calculated color profile, clearly exhibiting a U-shape, is shown in the bottom panel.

\section{Results}

\begin{figure*}
        \centering
        \includegraphics[width=0.9\textwidth]{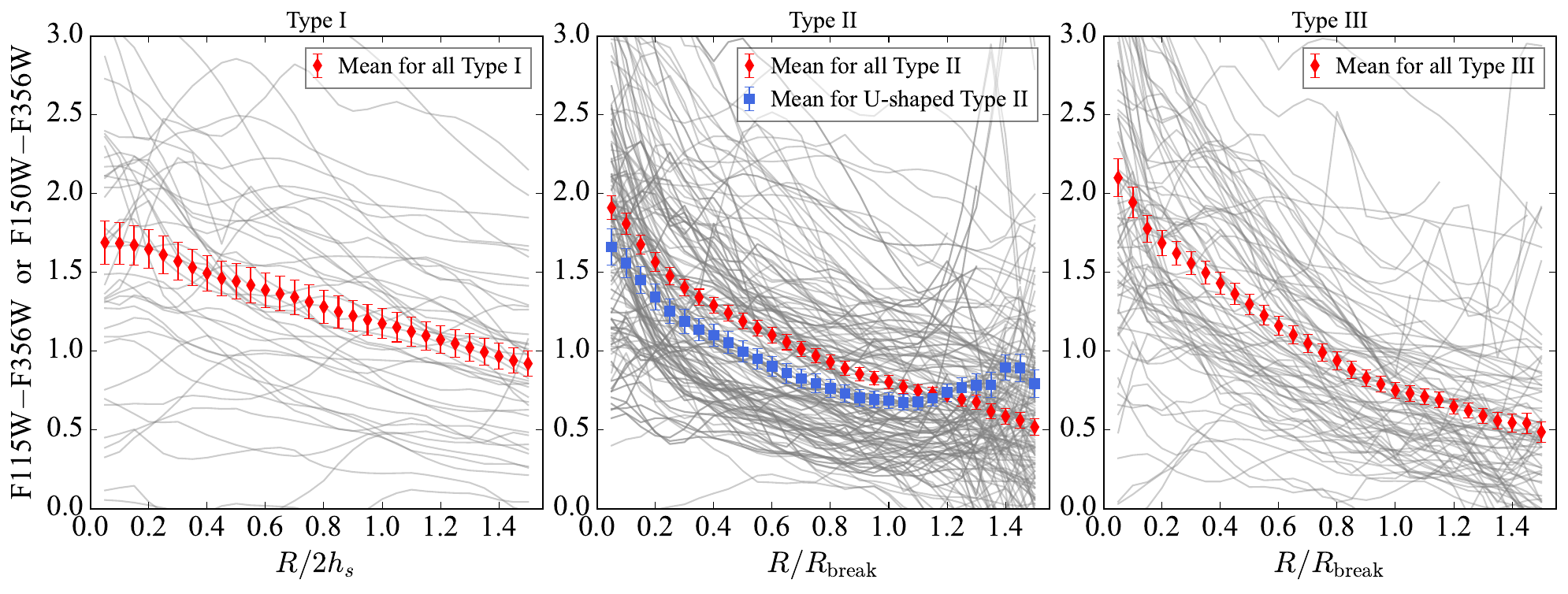}
        \caption{Color profiles of the galaxies at $z=1$--3 with JWST/NIRCam. The left, middle, and right panels show the results for Type~I, II, and III disks, respectively. The color index is chosen as the best proxy to the rest-frame $B-Y$ color. The radii are scaled to $2\times h_s$  for Type~I disks ---where $h_s$ is the scale
length---, and to the break radius $R_{\rm break}$ for Types~II and III. Gray curves are individual color profiles. Large red diamonds are the mean color profile for each subsample, and the error bars give the error. The large blue squares in the middle panel show the average color profile for U-shaped Type~II disks, with a minimum color of 0.67\,dex. 
        }
        \label{cp}
\end{figure*}

\begin{figure}
        \centering
        \includegraphics[width=0.45\textwidth]{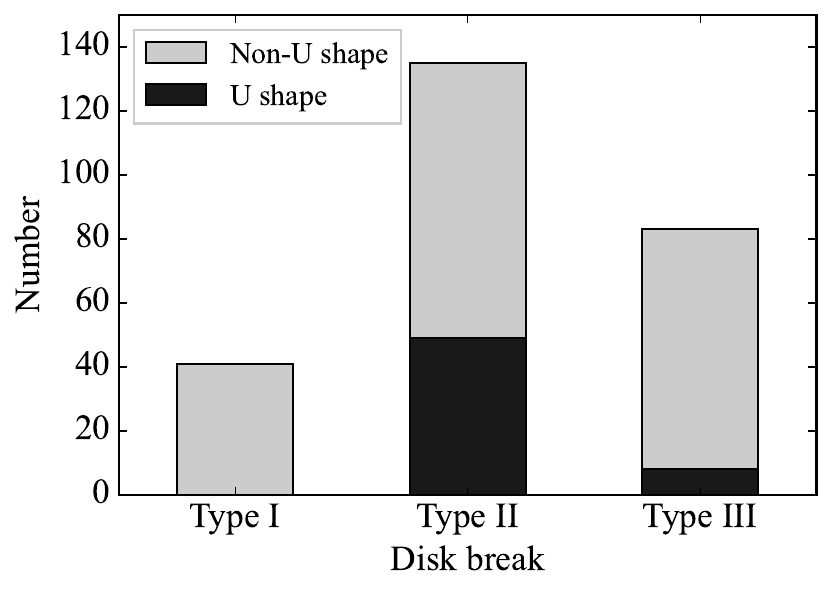}
        \caption{Distribution of disk galaxies by type and color profile shape. The black portion of each bar represents the number of galaxies with a U-shaped color profile, while the gray portion indicates those with a non-U shape. Specifically, 36\% of Type~II disks and 9\% of Type~III disks exhibit a U-shaped color profile. 
        }
        \label{distribution}
\end{figure}

\begin{figure}
        \centering 
        \includegraphics[width=0.45\textwidth]{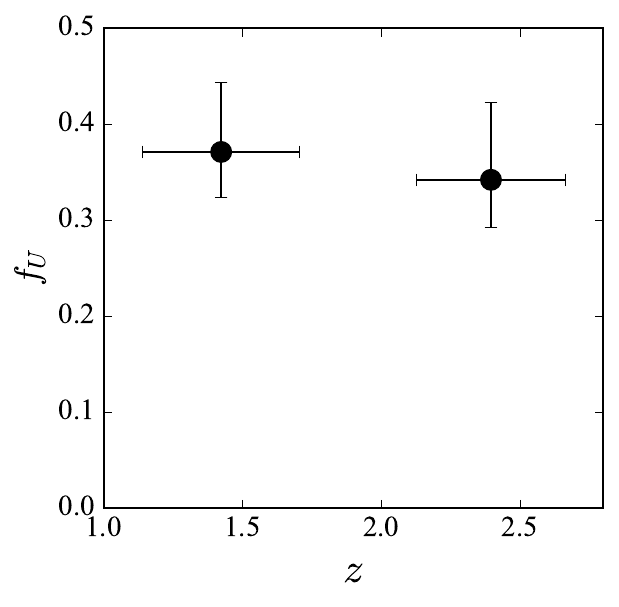}
        \caption{Fraction of Type II disks exhibiting U-shaped color profiles as a function of redshift $z$. The data points represent the observed fractions in two redshift bins: $37.1^{+4.8\%}_{-5.0\%}$ for $z=1$--2 and $34.2^{+7.2\%}_{-8.0\%}$ for $z=2$--3, with error bars reflecting the associated uncertainties.
        }
        \label{evo}
\end{figure}

Figure~\ref{cp} plots the color profiles of Type~I, II, and III disks in the left, middle, and right panels, respectively. The gray curves represent individual profiles. The variation in color is primarily due to stellar mass \citep{Azzollini2008_color, Bakos2008}. The large red diamonds indicate the average color profiles for each type. The average color profiles exhibit a systematic decrease with increasing radius, regardless of disk type. In particular, the average color profiles of all the high-redshift Type~II disks in the sample do not show the U-shape. This result differs significantly from the average color profiles observed at redshift $z<1$, where a U-shape is evident.

Previous analyses of disk galaxies at $z\approx0$ from the Sloan Digital Sky Survey and disk galaxies at $z=0.1$--1.1 from HST ACS imaging both revealed that the average color profiles of all the Type~II disks in their sample exhibit a U-shape near the disk break \citep{Bakos2008, Marino2016, Azzollini2008_color}. The apparent reason is that most of the low- and intermediate-redshift type~II disks have the U-shape.

We analyzed the color gradients to determine the shape of each color profile (Sect.~\ref{OM}) and present the distribution of disk galaxies by type and color profile shape in Fig.~\ref{distribution} to show how many of our high-redshift disks exhibit a U-shape. We find that 36\% (49 out of 135) of Type~II disks exhibit a U-shape at or near the disk break. The ratio of the break position to the U-shape position yields a mean value of $0.97$ with a scatter of $0.13$. The fraction of U-shaped color profiles in Type~II disks is significantly lower than the 70--90\% observed at redshift $z\lesssim 1$. 
In contrast, none and only 9\% (8 out of 83) of Type~I and Type III disks, respectively, exhibit a U-shaped profile. As a result, the average color profile of both Type~I and Type~III disks also lacks a U-shape, which is consistent with previous studies of galaxies at low and intermediate redshifts \citep{Azzollini2008_color, Bakos2008}.

To better illustrate the U-shape in high-redshift disks, we plot the average color profile for the 49 Type~II disks with a detected U-shape as blue squares in the middle panel of Fig.~\ref{cp}, where a clear U-shape is evident near the disk break. The average U-shape color profile exhibits a minimum value of 0.62\,dex. Additionally, we separate our galaxies into two redshift bins, $z=1$--2 and $z=2$--3, yielding fractions of U-shaped color profiles in Type~II disks of $37.1^{+4.8\%}_{-5.0\%}$ and 34.2$^{+7.2\%}_{-8.0\%}$, respectively. The uncertainties are calculated using the Wilson interval. These fractions do not show significant evolution across the redshift range considered.

\section{Implication}

Several models have been proposed to explain the different types of disk profiles. Type~I disks may form through the collapse of a uniformly rotating protogalactic cloud \citep{Freeman1970, Fall1980} or via viscous redistribution \citep[e.g.,][]{Ferguson2001}. Consequently, a U-shaped color profile is not expected in these disks.

Type~II profiles are genuine features of underlying older stellar disks \citep{Munoz-Mateos2013, Laine_2014}. The models proposed to explain the occurrence of the Type~II profile in stellar disks can be categorized into three branches: (1) protogalactic cloud collapse with angular-momentum conservation and a sharp cut-off in angular momentum \citep{vanderKruit1987}; (2) a star-formation threshold \citep{Kennicutt1989, Schaye2004, Elmegreen2006}; and (3) radial star migration \citep[e.g.,][]{Debattista2006, Bournaud2007, Roskar2008, Minchev2012, Martinez-Bautista2021}.

The collapse scenario is embedded within the inside-out galaxy-formation process \citep[e.g.,][]{Guo2011}. In this scenario, a galaxy forms in a cold dark matter halo and continues to accrete cold gas from the circumgalactic medium. The gas accreted initially has lower angular momentum, causing it to settle in the inner regions, while gas accreted later has higher angular momentum, leading it to fall into the outer regions. As a result, a sharp cut-off in the angular momentum of the accreted gas cloud can cause a truncation in the disk. However, this collapse scenario predicts a break radius at approximately 4.5 times the disk scale length, which is significantly larger than the observed break radii of 2--2.5 times the scale length \citep{Elmegreen2006}, making it an unlikely driver of the Type~II disk break. The star-formation threshold scenario suggests a suppression of star-formation activity in the outer region, where the molecular gas surface density falls below a threshold value, leading to a truncation in the disk. However, this scenario fails to produce a sufficiently large disk with an exponential profile extending into the outer region \citep{Schaye2004}, and thus does not fully account for the observed disk break. In addition to the star-formation threshold, a complementary mechanism, which relaxes the assumption that most stars currently in the outer disk were formed in situ, is also necessary.  Radial star migration, which relocates stars from the inner to outer disk, plays a critical role in making the outer down-bending disk.

Radial star migration can be driven by several mechanisms. \cite{Sellwood2002} proposed that transient spiral arms cause a dynamical churning effect, resulting in radial migration. \cite{Roskar2008} suggested that migration is primarily driven by the scattering effects of the bar and/or spiral corotation resonance, a secular process that unfolds over a few gigayears (see also \citealt{Debattista2006} and \citealt{Minchev2012}). \cite{Khoperskov2020} and \cite{Haywood2024} propose that bar structures can induce more rapid migration. Their results indicate that young stars trapped by resonances may be displaced outward within $\sim$1\,Gyr as the bar slows down. Additionally, gravitational torques from clumps, as demonstrated by \cite{Bournaud2007}, can effectively fling stars outward into the outer disk. This process is highly rapid, with timescales of 300--500\,Myr. Despite these complexities, the radial migration is crucial to the formation of Type~II disks. \cite{Roskar2008} concluded that Type~II disks form through the interplay between secular radial star migration and a star-formation cut-off in the outer disk, predicting an age profile with a U-shape at the break radius. The U-shape arises because secular migration moves old stars from the inner to the outer disk, while star formation in the outer regions has ceased. As young massive stars in the inner regions evolve rapidly and cease their main sequence phase before the secular radial migration takes effect, the stars transported outward are typically older. This scenario likely explains the 36\% of Type~II disks exhibiting U-shaped color profiles in our sample.  We note that while variations in radial dust thickness may contribute to bluer colors at larger radii, they do not explain the U-shaped color profile or the break in the surface-brightness profile.

However, 64\% of Type~II disks do not show U-shaped color profiles and thus  cannot be explained by the scenario involving secular radial migration. This is not entirely surprising, because secular processes require several gigayears to operate, and, additionally, the stellar structures that drive secular evolution, such as bars and spirals, might be only just forming at cosmic noon 
\citep[$z\approx1$--3; e.g.,][]{Kraljic2012, ElmegreenElmegreen2014}. These structures may not have had sufficient time for secular evolution to significantly impact the disk. The faster migration mechanism associated with de-acceleration of the bar, as proposed by \cite{Khoperskov2020} and \cite{Haywood2024}, may contribute to the formation of these disks. As this process is relatively fast, stars —including young stars— can be transferred outward within 1\,Gyr, potentially forming a break in the surface-brightness profile without producing a U-shape color profile. The absence of U-shaped color profiles in some Type~II disks could also be explained by radial migration driven by violent clump instabilities at high redshift, which can efficiently move stars outward on much shorter timescales \citep{Bournaud2007, Bournaud2016, Wu2020}.  A clump induces a kinematic response in the disk, creating over-densities that exchange angular momentum with the clump itself \citep{Bournaud2010}. The gravity torques in clump instabilities are 10--20 times stronger than in secular processes, enabling efficient angular-momentum transfer \citep{Bournaud2011}. As clumps lose angular momentum and move inward, stars in the disk gain angular momentum and migrate outward \citep{Bournaud2007}. Clumps in high-redshift galaxies are now detected in near-infrared using JWST \citep{Boris2024a, Boris2024b}. Although clump-instability models \citep[e.g.,][]{Bournaud2007} do not provide any prediction of the age distribution of stars across the radial range, we infer that, in addition to old stars, young massive stars born in the inner regions could be transported to the outer regions in such a rapid process. If the disk break is driven by a combination of a star-formation threshold and rapid radial migration via violent clump instabilities, then Type~II disks without U-shaped color profiles could form under these conditions.

For Type~III galaxies, the formation mechanisms are less well understood. Minor mergers, gas accretion, and interactions with the intergalactic or circumgalactic environment may all contribute to their formation \citep{Bekki1998, Younger2007, Minchev2012, Watkins2019, Peters2017, Wang2018}. \citet{Bakos2012} suggests that some previously identified Type~III profiles might be contaminated by stellar halos, arguing that true Type III disks may not exist. Nevertheless, combining these mechanisms, the truncation of recent star formation in the outer disk could account for the few observed Type~III disks with U-shaped color profiles.

\section{Conclusion}

We analyzed the approximate rest-frame $B-Y$ color profiles of 223 disk galaxies at redshifts of $z=1$--3 observed by JWST, focusing on their relation to disk breaks. We find that 36\% of Type~II disks exhibit U-shaped color profiles with a minimum at or near the disk break, suggesting the influence of secular radial migration of old stars combined with a star formation threshold in these systems. The fraction of U-shaped color in Type~II disks remains almost consistent across the redshift range $z=1$--3, but is significantly lower than that observed in the local Universe. These results indicate that while secular processes are important, they may not have had sufficient time  ---a  few gigayears--- to fully shape disk galaxies at high redshifts. In contrast, Type~II disks without U-shaped color profiles may suggest rapid rather than secular radial star migration. This may partly align with the scenario proposed by \cite{Khoperskov2020} and \cite{Haywood2024}, wherein stars are captured by resonances and are subsequently  transferred outward within $\sim$1\,Gyr. Additionally, rapid star migration is also potentially driven by violent clump instabilities \citep{Roskar2008}, which can transport both young and old stars outward on timescales of 300--500\,Myr, forming the secondary down-bending outer disk. These findings may offer critical insights into the formation and evolution of disk structures in the early Universe.

\begin{acknowledgements}
We are grateful to the anonymous referee for helpful feedback.
TF and XL acknowledge support from the National Science Foundation of China (11890692, 12133008, 12221003) and the China Manned Space Project (CMSCSST-2021-A04). Kavli IPMU is supported by World Premier International Research Center Initiative (WPI), MEXT, Japan.

\end{acknowledgements}

\bibliographystyle{aa}

\begin{thebibliography}{46}
\expandafter\ifx\csname natexlab\endcsname\relax\def\natexlab#1{#1}\fi

\bibitem[{{Azzollini} {et~al.}(2008{\natexlab{a}}){Azzollini}, {Trujillo}, \&
  {Beckman}}]{Azzollini2008_color}
{Azzollini}, R., {Trujillo}, I., \& {Beckman}, J.~E. 2008{\natexlab{a}}, \apjl,
  679, L69

\bibitem[{{Azzollini} {et~al.}(2008{\natexlab{b}}){Azzollini}, {Trujillo}, \&
  {Beckman}}]{Azzollini2008}
{Azzollini}, R., {Trujillo}, I., \& {Beckman}, J.~E. 2008{\natexlab{b}}, \apj,
  684, 1026

\bibitem[{{Bakos} \& {Trujillo}(2012)}]{Bakos2012}
{Bakos}, J. \& {Trujillo}, I. 2012, arXiv e-prints, arXiv:1204.3082

\bibitem[{{Bakos} {et~al.}(2008){Bakos}, {Trujillo}, \& {Pohlen}}]{Bakos2008}
{Bakos}, J., {Trujillo}, I., \& {Pohlen}, M. 2008, \apjl, 683, L103

\bibitem[{{Bekki}(1998)}]{Bekki1998}
{Bekki}, K. 1998, \apjl, 502, L133

\bibitem[{{Bournaud}(2010)}]{Bournaud2010}
{Bournaud}, F. 2010, in Astronomical Society of the Pacific Conference Series,
  Vol. 423, Galaxy Wars: Stellar Populations and Star Formation in Interacting
  Galaxies, ed. B.~{Smith}, J.~{Higdon}, S.~{Higdon}, \& N.~{Bastian}, 177

\bibitem[{{Bournaud}(2016)}]{Bournaud2016}
{Bournaud}, F. 2016, in Astrophysics and Space Science Library, Vol. 418,
  Galactic Bulges, ed. E.~{Laurikainen}, R.~{Peletier}, \& D.~{Gadotti}, 355

\bibitem[{{Bournaud} {et~al.}(2011){Bournaud}, {Dekel}, {Teyssier}, {Cacciato},
  {Daddi}, {Juneau}, \& {Shankar}}]{Bournaud2011}
{Bournaud}, F., {Dekel}, A., {Teyssier}, R., {et~al.} 2011, \apjl, 741, L33

\bibitem[{{Bournaud} {et~al.}(2007){Bournaud}, {Elmegreen}, \&
  {Elmegreen}}]{Bournaud2007}
{Bournaud}, F., {Elmegreen}, B.~G., \& {Elmegreen}, D.~M. 2007, \apj, 670, 237

\bibitem[{{Debattista} {et~al.}(2006){Debattista}, {Mayer}, {Carollo}, {Moore},
  {Wadsley}, \& {Quinn}}]{Debattista2006}
{Debattista}, V.~P., {Mayer}, L., {Carollo}, C.~M., {et~al.} 2006, \apj, 645,
  209

\bibitem[{{Elmegreen} \& {Hunter}(2006)}]{Elmegreen2006}
{Elmegreen}, B.~G. \& {Hunter}, D.~A. 2006, \apj, 636, 712

\bibitem[{{Elmegreen} \& {Elmegreen}(2014)}]{ElmegreenElmegreen2014}
{Elmegreen}, D.~M. \& {Elmegreen}, B.~G. 2014, \apj, 781, 11

\bibitem[{{Erwin} {et~al.}(2008){Erwin}, {Pohlen}, \& {Beckman}}]{Erwin2008}
{Erwin}, P., {Pohlen}, M., \& {Beckman}, J.~E. 2008, \aj, 135, 20

\bibitem[{{Fall} \& {Efstathiou}(1980)}]{Fall1980}
{Fall}, S.~M. \& {Efstathiou}, G. 1980, \mnras, 193, 189

\bibitem[{{Ferguson} \& {Clarke}(2001)}]{Ferguson2001}
{Ferguson}, A.~M.~N. \& {Clarke}, C.~J. 2001, \mnras, 325, 781

\bibitem[{{Finkelstein} {et~al.}(2022){Finkelstein}, {Bagley}, {Haro},
  {Dickinson}, {Ferguson}, {Kartaltepe}, {Papovich}, {Burgarella}, {Kocevski},
  {Huertas-Company}, {Iyer}, {Koekemoer}, {Larson}, {P{\'e}rez-Gonz{\'a}lez},
  {Rose}, {Tacchella}, {Wilkins}, {Chworowsky}, {Medrano}, {Morales},
  {Somerville}, {Yung}, {Fontana}, {Giavalisco}, {Grazian}, {Grogin}, {Kewley},
  {Kirkpatrick}, {Kurczynski}, {Lotz}, {Pentericci}, {Pirzkal}, {Ravindranath},
  {Ryan}, {Trump}, {Yang}, {Almaini}, {Amor{\'\i}n}, {Annunziatella},
  {Backhaus}, {Barro}, {Behroozi}, {Bell}, {Bhatawdekar}, {Bisigello}, {Bromm},
  {Buat}, {Buitrago}, {Calabr{\`o}}, {Casey}, {Castellano}, {Ch{\'a}vez Ortiz},
  {Ciesla}, {Cleri}, {Cohen}, {Cole}, {Cooke}, {Cooper}, {Cooray}, {Costantin},
  {Cox}, {Croton}, {Daddi}, {Dav{\'e}}, {de La Vega}, {Dekel}, {Elbaz},
  {Estrada-Carpenter}, {Faber}, {Fern{\'a}ndez}, {Finkelstein}, {Freundlich},
  {Fujimoto}, {Garc{\'\i}a-Argum{\'a}nez}, {Gardner}, {Gawiser},
  {G{\'o}mez-Guijarro}, {Guo}, {Hamblin}, {Hamilton}, {Hathi}, {Holwerda},
  {Hirschmann}, {Hutchison}, {Jaskot}, {Jha}, {Jogee}, {Juneau}, {Jung},
  {Kassin}, {Bail}, {Leung}, {Lucas}, {Magnelli}, {Mantha}, {Matharu},
  {McGrath}, {McIntosh}, {Merlin}, {Mobasher}, {Newman}, {Nicholls}, {Pandya},
  {Rafelski}, {Ronayne}, {Santini}, {Seill{\'e}}, {Shah}, {Shen}, {Simons},
  {Snyder}, {Stanway}, {Straughn}, {Teplitz}, {Vanderhoof}, {Vega-Ferrero},
  {Wang}, {Weiner}, {Willmer}, {Wuyts}, {Zavala}, \& {Ceers
  Team}}]{Finkelstein:2022}
{Finkelstein}, S.~L., {Bagley}, M.~B., {Haro}, P.~A., {et~al.} 2022, \apjl,
  940, L55

\bibitem[{{Freeman}(1970)}]{Freeman1970}
{Freeman}, K.~C. 1970, \apj, 160, 811

\bibitem[{{Guo} {et~al.}(2011){Guo}, {White}, {Boylan-Kolchin}, {De Lucia},
  {Kauffmann}, {Lemson}, {Li}, {Springel}, \& {Weinmann}}]{Guo2011}
{Guo}, Q., {White}, S., {Boylan-Kolchin}, M., {et~al.} 2011, \mnras, 413, 101

\bibitem[{{Guti{\'e}rrez} {et~al.}(2011){Guti{\'e}rrez}, {Erwin}, {Aladro}, \&
  {Beckman}}]{Gutirrez2011}
{Guti{\'e}rrez}, L., {Erwin}, P., {Aladro}, R., \& {Beckman}, J.~E. 2011, \aj,
  142, 145

\bibitem[{{Haywood} {et~al.}(2024){Haywood}, {Khoperskov}, {Cerqui}, {Di
  Matteo}, {Katz}, \& {Snaith}}]{Haywood2024}
{Haywood}, M., {Khoperskov}, S., {Cerqui}, V., {et~al.} 2024, \aap, 690, A147

\bibitem[{{Kalita} {et~al.}(2024{\natexlab{a}}){Kalita}, {Silverman}, {Daddi},
  {Bottrell}, {Ho}, {Ding}, \& {Yang}}]{Boris2024a}
{Kalita}, B.~S., {Silverman}, J.~D., {Daddi}, E., {et~al.} 2024{\natexlab{a}},
  \apj, 960, 25

\bibitem[{{Kalita} {et~al.}(2024{\natexlab{b}}){Kalita}, {Silverman}, {Daddi},
  {Mercier}, {Ho}, \& {Ding}}]{Boris2024b}
{Kalita}, B.~S., {Silverman}, J.~D., {Daddi}, E., {et~al.} 2024{\natexlab{b}},
  arXiv e-prints, arXiv:2402.02679

\bibitem[{{Kennicutt}(1989)}]{Kennicutt1989}
{Kennicutt}, Robert~C., J. 1989, \apj, 344, 685

\bibitem[{{Khoperskov} {et~al.}(2020){Khoperskov}, {Di Matteo}, {Haywood},
  {G{\'o}mez}, \& {Snaith}}]{Khoperskov2020}
{Khoperskov}, S., {Di Matteo}, P., {Haywood}, M., {G{\'o}mez}, A., \& {Snaith},
  O.~N. 2020, \aap, 638, A144

\bibitem[{{Kraljic} {et~al.}(2012){Kraljic}, {Bournaud}, \&
  {Martig}}]{Kraljic2012}
{Kraljic}, K., {Bournaud}, F., \& {Martig}, M. 2012, \apj, 757, 60

\bibitem[{{Laine} {et~al.}(2014){Laine}, {Laurikainen}, {Salo}, {Comer{\'o}n},
  {Buta}, {Zaritsky}, {Athanassoula}, {Bosma}, {Mu{\~n}oz-Mateos}, {Gadotti},
  {Hinz}, {Erroz-Ferrer}, {Gil de Paz}, {Kim}, {Men{\'e}ndez-Delmestre},
  {Mizusawa}, {Regan}, {Seibert}, \& {Sheth}}]{Laine_2014}
{Laine}, J., {Laurikainen}, E., {Salo}, H., {et~al.} 2014, \mnras, 441, 1992

\bibitem[{{Liang} {et~al.}(2024){Liang}, {Yu}, {Fang}, \& {Ho}}]{Liang2024}
{Liang}, X., {Yu}, S.-Y., {Fang}, T., \& {Ho}, L.~C. 2024, \aap, 688, A158

\bibitem[{{Marino} {et~al.}(2016){Marino}, {Gil de Paz}, {S{\'a}nchez},
  {S{\'a}nchez-Bl{\'a}zquez}, {Cardiel}, {Castillo-Morales}, {Pascual},
  {V{\'\i}lchez}, {Kehrig}, {Moll{\'a}}, {Mendez-Abreu},
  {Catal{\'a}n-Torrecilla}, {Florido}, {Perez}, {Ruiz-Lara}, {Ellis},
  {L{\'o}pez-S{\'a}nchez}, {Gonz{\'a}lez Delgado}, {de Lorenzo-C{\'a}ceres},
  {Garc{\'\i}a-Benito}, {Galbany}, {Zibetti}, {Cortijo}, {Kalinova}, {Mast},
  {Iglesias-P{\'a}ramo}, {Papaderos}, {Walcher}, \&
  {Bland-Hawthorn}}]{Marino2016}
{Marino}, R.~A., {Gil de Paz}, A., {S{\'a}nchez}, S.~F., {et~al.} 2016, \aap,
  585, A47

\bibitem[{{Mart{\'\i}nez-Bautista} {et~al.}(2021){Mart{\'\i}nez-Bautista},
  {Vel{\'a}zquez}, {P{\'e}rez-Villegas}, \& {Moreno}}]{Martinez-Bautista2021}
{Mart{\'\i}nez-Bautista}, G., {Vel{\'a}zquez}, H., {P{\'e}rez-Villegas}, A., \&
  {Moreno}, E. 2021, \mnras, 504, 5919

\bibitem[{{Minchev} {et~al.}(2012){Minchev}, {Famaey}, {Quillen}, {Di Matteo},
  {Combes}, {Vlaji{\'c}}, {Erwin}, \& {Bland-Hawthorn}}]{Minchev2012}
{Minchev}, I., {Famaey}, B., {Quillen}, A.~C., {et~al.} 2012, \aap, 548, A126

\bibitem[{{Mu{\~n}oz-Mateos} {et~al.}(2013){Mu{\~n}oz-Mateos}, {Sheth}, {Gil de
  Paz}, {Meidt}, {Athanassoula}, {Bosma}, {Comer{\'o}n}, {Elmegreen},
  {Elmegreen}, {Erroz-Ferrer}, {Gadotti}, {Hinz}, {Ho}, {Holwerda}, {Jarrett},
  {Kim}, {Knapen}, {Laine}, {Laurikainen}, {Madore}, {Menendez-Delmestre},
  {Mizusawa}, {Regan}, {Salo}, {Schinnerer}, {Seibert}, {Skibba}, \&
  {Zaritsky}}]{Munoz-Mateos2013}
{Mu{\~n}oz-Mateos}, J.~C., {Sheth}, K., {Gil de Paz}, A., {et~al.} 2013, \apj,
  771, 59

\bibitem[{{Peters} {et~al.}(2017){Peters}, {van der Kruit}, {Knapen},
  {Trujillo}, {Fliri}, {Cisternas}, \& {Kelvin}}]{Peters2017}
{Peters}, S.~P.~C., {van der Kruit}, P.~C., {Knapen}, J.~H., {et~al.} 2017,
  \mnras, 470, 427

\bibitem[{{Pohlen} \& {Trujillo}(2006)}]{pohlen2006}
{Pohlen}, M. \& {Trujillo}, I. 2006, \aap, 454, 759

\bibitem[{{Rosas-Guevara} {et~al.}(2022){Rosas-Guevara}, {Bonoli}, {Dotti},
  {Izquierdo-Villalba}, {Lupi}, {Zana}, {Bonetti}, {Nelson}, {Springel},
  {Hernquist}, \& {Vogelsberger}}]{Rosas-Guevara2022}
{Rosas-Guevara}, Y., {Bonoli}, S., {Dotti}, M., {et~al.} 2022, \mnras, 512,
  5339

\bibitem[{{Ro{\v{s}}kar} {et~al.}(2008){Ro{\v{s}}kar}, {Debattista}, {Stinson},
  {Quinn}, {Kaufmann}, \& {Wadsley}}]{Roskar2008}
{Ro{\v{s}}kar}, R., {Debattista}, V.~P., {Stinson}, G.~S., {et~al.} 2008,
  \apjl, 675, L65

\bibitem[{{Schaye}(2004)}]{Schaye2004}
{Schaye}, J. 2004, \apj, 609, 667

\bibitem[{{Sellwood} \& {Binney}(2002)}]{Sellwood2002}
{Sellwood}, J.~A. \& {Binney}, J.~J. 2002, \mnras, 336, 785

\bibitem[{{van der Kruit}(1979)}]{vanderKruit1979}
{van der Kruit}, P.~C. 1979, \aaps, 38, 15

\bibitem[{{van der Kruit}(1987)}]{vanderKruit1987}
{van der Kruit}, P.~C. 1987, \aap, 173, 59

\bibitem[{{Wang} {et~al.}(2018){Wang}, {Zheng}, {D'Souza}, {Mo}, {J{\'o}zsa},
  {Li}, {Kamphuis}, {Catinella}, {Shao}, {Lagos}, {Du}, \& {Pan}}]{Wang2018}
{Wang}, J., {Zheng}, Z., {D'Souza}, R., {et~al.} 2018, \mnras, 479, 4292

\bibitem[{{Watkins} {et~al.}(2019){Watkins}, {Laine}, {Comer{\'o}n}, {Janz}, \&
  {Salo}}]{Watkins2019}
{Watkins}, A.~E., {Laine}, J., {Comer{\'o}n}, S., {Janz}, J., \& {Salo}, H.
  2019, \aap, 625, A36

\bibitem[{{Wu} {et~al.}(2020){Wu}, {Struck}, {D'Onghia}, \&
  {Elmegreen}}]{Wu2020}
{Wu}, J., {Struck}, C., {D'Onghia}, E., \& {Elmegreen}, B.~G. 2020, \mnras,
  499, 2672

\bibitem[{{Xu} \& {Yu}(2024)}]{XuYu2024}
{Xu}, D. \& {Yu}, S.-Y. 2024, \aap, 682, L17

\bibitem[{{Younger} {et~al.}(2007){Younger}, {Cox}, {Seth}, \&
  {Hernquist}}]{Younger2007}
{Younger}, J.~D., {Cox}, T.~J., {Seth}, A.~C., \& {Hernquist}, L. 2007, \apj,
  670, 269

\bibitem[{{Yu} {et~al.}(2023){Yu}, {Cheng}, {Pan}, {Sun}, \& {Li}}]{Yu2023}
{Yu}, S.-Y., {Cheng}, C., {Pan}, Y., {Sun}, F., \& {Li}, Y.~A. 2023, \aap, 676,
  A74

\bibitem[{{Zheng} {et~al.}(2015){Zheng}, {Thilker}, {Heckman}, {Meurer},
  {Burgett}, {Chambers}, {Huber}, {Kaiser}, {Magnier}, {Metcalfe}, {Price},
  {Tonry}, {Wainscoat}, \& {Waters}}]{Zheng2015}
{Zheng}, Z., {Thilker}, D.~A., {Heckman}, T.~M., {et~al.} 2015, \apj, 800, 120

\end{thebibliography}


\end{document}